\documentclass[a4paper]{JACoW-GSI}
\usepackage{graphicx}
\usepackage{url}
\usepackage{epsfig}

\def\bea{\begin{eqnarray}}
\def\eea{\end{eqnarray}}
\def\be{\begin{equation}}
\def\ee{\end{equation}}

\def\lQ{\Lambda_{\rm QCD}}

\def\als{\alpha_{\rm s}}

\def\siml{{\ \lower-1.2pt\vbox{\hbox{\rlap{$<$}\lower6pt\vbox{\hbox{$\sim$}}}}\ }}     
\def\simg{{\
    \lower-1.2pt\vbox{\hbox{\rlap{$>$}\lower6pt\vbox{\hbox{$\sim$}}}}\ }} 
\newcommand{\nn}{\nonumber}
\newcommand{\lh}{\rm l.h.}

\begin{document}
\title{\hfill {\small TUM-EFT 5/09} \\ Quarkonia:  a theoretical frame\\ \vspace{-0.5cm} }
%\title{Quarkonia:  a theoretical frame\\ \vspace{-0.5cm} }
\author{Antonio Vairo}
\affil{Physik Department, Technische Universit\"at M\"unchen,
James-Frank-Str. 1, 85748 Garching, Germany}

\maketitle

\section{Why to study quarkonia}
Quarkonia, i.e. bound states made of a heavy quark and a heavy antiquark
(like charmonia, bottomonia, ...) 
are systems where low-energy QCD may be studied in a systematic way
(e.g. one may address issues like large-order perturbation theory, 
non-perturbative matrix elements, QCD
vacuum, exotica, confinement, deconfinement, ... ).
This is because $M \gg p \gg E$, where $M$ is the heavy-quark mass, 
$p$ the momentum transfer and $E$ the binding energy of the bound state, and because 
$M \gg \lQ$, the scale of non-perturbative QCD.

\begin{itemize}
\item[(1)]{$M \gg p \gg E$ implies that quarkonia are non-relativistic and characterized by the
hierarchy of scales typical of a non-relativistic bound state:
\bea
M \gg p \sim 1/r \sim Mv \gg E \sim M v^2,
\eea 
where $r$ is the typical distance between the heavy quark and the heavy antiquark 
and $v \ll 1$ is the typical heavy-quark velocity.
Systematic expansions in $v$ may be implemented at the
Lagrangian level by constructing suitable effective field theories (EFTs):
expanding QCD in $p/M$ and $E/M$ leads to NRQCD \cite{Caswell:1985ui};
expanding NRQCD in $E\,r$ leads to pNRQCD \cite{Pineda:1997bj}.
The hierarchy of non-relativistic scales makes the very
difference of quarkonia from heavy-light mesons, which are characterized
by just two scales: $M$ and $\lQ$ \cite{Neubert:1993mb}.
For a review of non-relativistic effective field theories we refer to 
\cite{Brambilla:2004jw}. Some historical background has been discussed in 
\cite{Vairo:2009rs}.
}
\item[(2)]{$M \gg \lQ$ implies $\als(M) \ll 1$: phenomena happening 
at the scale $M$ may be treated perturbatively.
We may further have small couplings if $Mv \gg \lQ$ and $Mv^2 \gg \lQ$, 
in which case  $\als(Mv) \ll 1$ and $\als(Mv^2)\ll 1$ respectively. 
This is likely to happen only for the lowest charmonium and bottomonium
states (see Fig.~\ref{figalphas}). 

\begin{figure}[h]
\makebox[0truecm]{\phantom b}
\put(20,0){\epsfxsize=6truecm \epsfbox{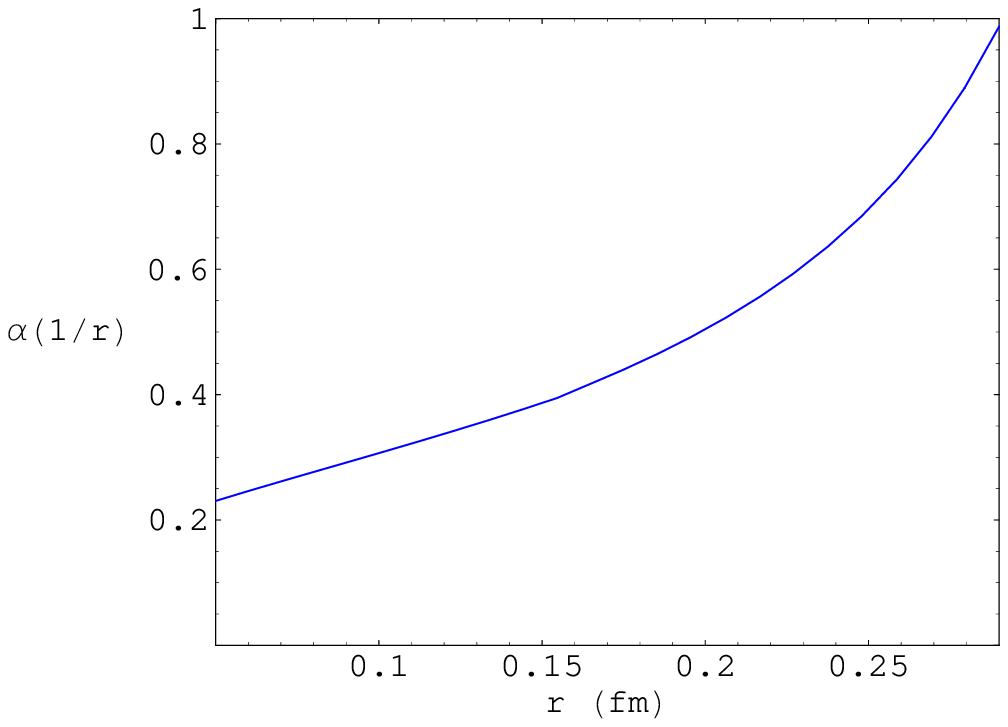}}
\put(115,35){\epsfxsize=0.1truecm \epsfbox{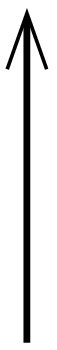}}
\put(110,22){ $\Upsilon$}
\put(165,35){\epsfxsize=0.1truecm \epsfbox{uparrow.eps}}
\put(145,22){ $J/\psi$, $\Upsilon'$}
\caption{$\als$ running at one loop and $\als(M_{J/\psi}v_{J/\psi})$ 
and $\als(M_{\Upsilon(1S)}v_{\Upsilon(1S)})$.}
\label{figalphas}
\end{figure}
}
\end{itemize}

\begin{figure}[h]
\makebox[0truecm]{\phantom b}
\put(25,2){\epsfxsize=6truecm \epsfbox{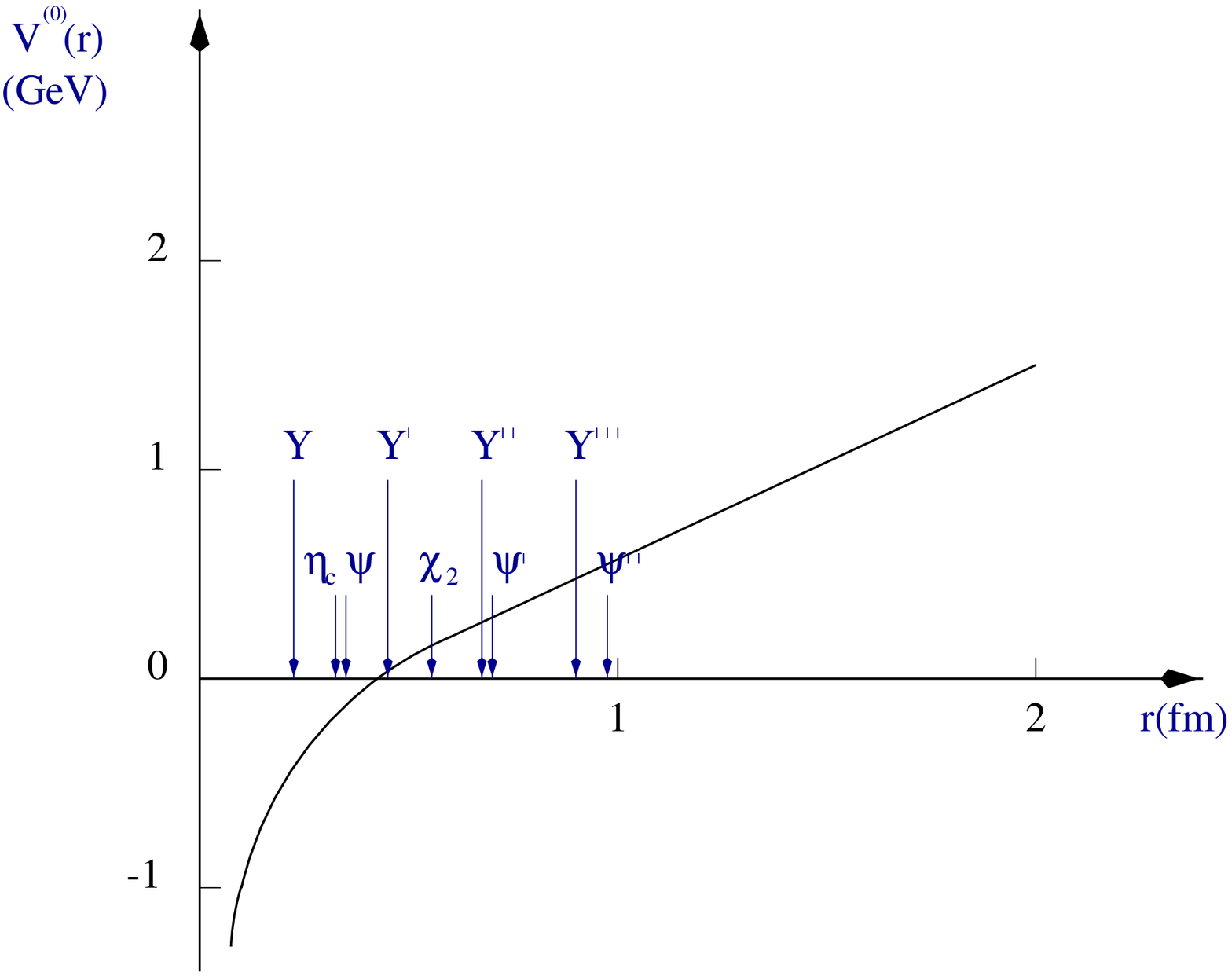}}
\put(70,2){\epsfxsize=1.5truecm \epsfbox{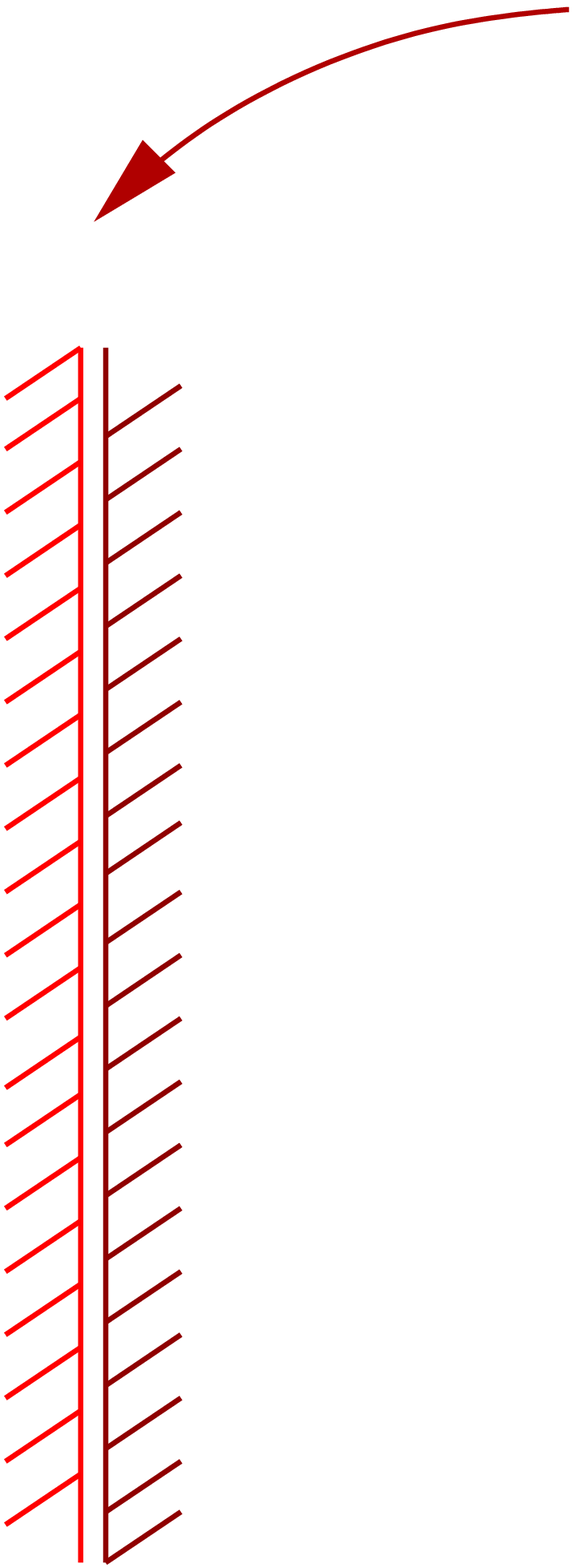}}
\put(115,115){$\lQ$}
\put(10,90){\it   Low-lying $Q\bar{Q}$}
\put(88,90){\it   High-lying $Q\bar{Q}$}
\caption{Static potential vs quarkonium radii taken from \cite{Godfrey:1985xj}.}
\label{figGI}
\end{figure}

It is precisely the rich structure of separated energy scales that makes
quarkonium an ideal probe of confinement and deconfinement. 
The different quarkonium radii are differently sensitive to the 
Coulombic and confining interaction (see Fig.~\ref{figGI}).
Hence, different quarkonia will dissociate in a medium at different
temperatures, providing a thermometer for the plasma \cite{Matsui:1986dk,Petreczky09}. 

\begin{figure}[h]
\makebox[0truecm]{\phantom b}
\put(20,0){\epsfxsize=6truecm \epsfbox{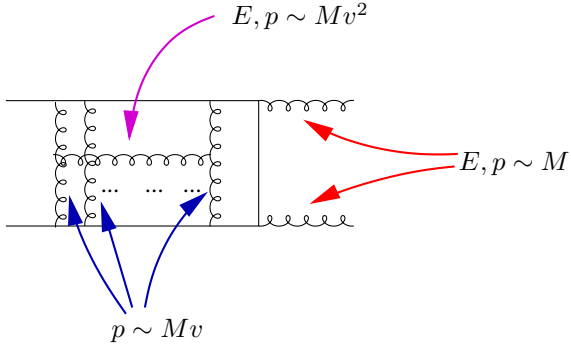}}
\put(60,-8){$p\sim Mv$}
\put(190,55){$E,p\sim M$}
\put(105,110){$E,p\sim Mv^2$}
\caption{Typical scales appearing in a quarkonium annihilation diagram.}
\label{figent}
\end{figure}

All quarkonium scales get entangled in a typical amplitude involving a quarkonium
observable, see Fig.~\ref{figent}. In particular, quarkonium annihilation 
and production happen at the scale $M$, quarkonium binding happens at the scale
$Mv$, which is the typical momentum exchanged inside the bound state, while 
very low-energy gluons and light quarks (also called ultrasoft degrees of freedom)  
live long enough that a bound state has time to form and, therefore, are sensitive to the 
scale $Mv^2$. Ultrasoft gluons are responsible for phenomena like the Lamb shift in QCD.

\section{Physics at the scale $M$}
\begin{figure}[h]
\makebox[0cm]{\phantom b}
\put(0,0){\epsfxsize=7truecm\epsffile{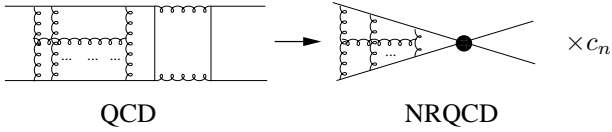}}
\put(36,-15){QCD}
\put(150,-15){NRQCD}
\put(210,12){$\times  c_n$}
\makebox[0cm]{\phantom b}
\caption{Matching of the diagram of Fig.~\ref{figent} to NRQCD.}
\label{fignrqcd}
\end{figure}

Quarkonium annihilation and production happen at the scale $M$. The suitable EFT is NRQCD, which 
follows from QCD by integrating out the scale $M$, see Fig.~\ref{fignrqcd}. As a consequence, 
the effective Lagrangian is organized as an expansion in $1/M$  and $\als(M)$: 
\be
{\cal L}_{\rm NRQCD}  = \sum_n \frac{c_n(\als(M),\mu)}{M^{n} } \times  O_n(\mu,Mv,Mv^2,...),
\ee
where $O_n$ are the operators of NRQCD that live at the low-energy scales $Mv$
and $Mv^2$, $\mu$ is the NRQCD factorization scale and $c_n$ are the Wilson
coefficients of the EFT that encode the contributions from the scale $M$ and are
non-analytic in $M$.

The NRQCD factorization formula for the quarkonium annihilation width (into light hadrons or photons or 
lepton pairs) reads:
\be
\Gamma_{H \hbox{ annihilation}} = 
\sum_n \frac{2 \, {\rm Im} \, c_n }{ M^{d_{O_n} - 4}} \langle H | O^{\rm 4-fermion}_n |H \rangle, 
\ee
where $d_{O_n}$ is the dimension of the four-fermion operator $O^{\rm 4-fermion}_n$.
Recently, substantial progress has been made in the evaluation of the factorization formula at order $v^7$
\cite{Brambilla:2006ph,Brambilla:2008zg}, in the lattice evaluation of the NRQCD matrix elements 
$\langle H | O^{\rm 4-fermion}_n |H \rangle$ \cite{Bodwin:2005gg} and in the data of many hadronic 
and electromagnetic decays \cite{Brambilla:2004wf}. As it was already discussed in \cite{Brambilla:2004wf}, 
the data are clearly sensitive to NLO corrections in the Wilson coefficients $c_n$ 
(and presumably also to relativistic corrections). For an updated list of
ratios of P-wave charmonium decay widths, see Tab.~\ref{tabdec}

\begin{table}
\begin{center}
\begin{tabular}{|c||c||c|c|}
\hline
Ratio &{ PDG09} &LO & NLO 
 \\
\hline
$\displaystyle \frac{\Gamma_{\chi_{c0}\to\gamma\gamma}}{ \Gamma_{\chi_{c2}\to\gamma\gamma}}$ 
& $\approx$ 4.9 &   3.75  & $\approx$ 5.43   \\
\hline
$\displaystyle \frac{\Gamma_{\chi_{c2}\to \lh} - \Gamma_{\chi_{c1}\to \lh}}{ \Gamma_{\chi_{c0}\to\gamma\gamma}}$ 
& $\approx$ 440 &  $\approx$ 347  & $\approx$ 383   \\
\hline
$\displaystyle \frac{\Gamma_{\chi_{c0}\to \lh} - \Gamma_{\chi_{c1}\to \lh}}{ \Gamma_{\chi_{c0}\to\gamma\gamma}}$ 
& $\approx$ 4000 &  $\approx$ 1300  & $\approx$ 2781   \\
\hline
$\displaystyle \frac{\Gamma_{\chi_{c0}\to \lh} - \Gamma_{\chi_{c2}\to \lh} 
}{ \Gamma_{\chi_{c2}\to \lh}  - \Gamma_{\chi_{c1}\to \lh }}$ 
& $\approx$ 8.0 &  2.75  & $\approx$ 6.63   \\
\hline
$\displaystyle \frac{\Gamma_{\chi_{c0}\to \lh}  - \Gamma_{\chi_{c1}\to \lh}  }
{ \Gamma_{\chi_{c2}\to \lh} - \Gamma_{\chi_{c1}\to \lh}}$ 
& $\approx$ 9.0 &  3.75  & $\approx$ 7.63   \\
\hline
\end{tabular}
\end{center}
\caption{Comparison of decay width ratios of $\chi_{cJ}$ from \cite{Amsler:2008zzb} 
(l.h. stands for light hadrons) with LO and NLO determinations (without 
corrections of relative order $v^2$, $m_c =$ 1.5 GeV and $\als(2m_c) =$ 0.245). }
\label{tabdec}
\end{table}

The high precision of data and matrix elements has been recently exploited to 
provide a new determination of $\als$ from 
$\Gamma_{\Upsilon(1S)\to \gamma \; \lh}/\Gamma_{\Upsilon(1S)\to \lh}$ \cite{Brambilla:2007cz}:
\be
\als (M_{\Upsilon(1S)})= 0.184^{+0.015}_{-0.014}, \qquad \als(M_Z)=0.119^{+0.006}_{-0.005}. 
\ee

Recent progress in quarkonium production in the framework of the NRQCD factorization 
have been discussed in \cite{Artoisenet09}.

\section{Physics at the scales $Mv$ and $Mv^2$} 
\begin{figure}
\makebox[0cm]{\phantom b}
\put(0,0){{\epsfxsize=8truecm\epsffile{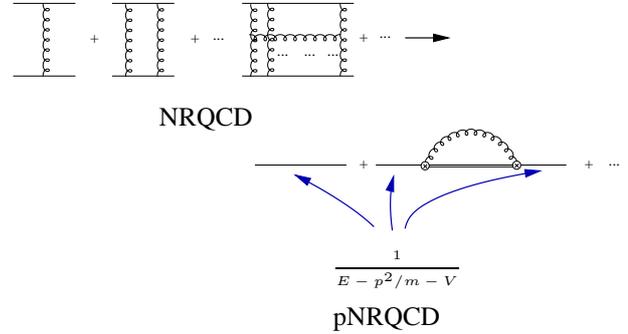}}}
\put(55,40){NRQCD}
\put(120,-35){pNRQCD}
\put(120,-15){\tiny $\displaystyle \frac{1}{E - p^2/m-V}$}
\makebox[0cm]{\phantom b}
\caption{Matching of the diagram of Fig.~\ref{fignrqcd} to pNRQCD.}
\label{figpnrqcd}
\end{figure}

Quarkonium formation happens at the scale $Mv$. 
The suitable EFT is pNRQCD \cite{Pineda:1997bj}, which 
follows from NRQCD by integrating out the scale $Mv$, see Fig.~\ref{figpnrqcd}. 
As a consequence, the effective Lagrangian is organized as an expansion in $1/M$  and $\als(M)$, 
inherited from NRQCD, and an expansion in $r$: 
\bea
&& {\cal L}_{\rm pNRQCD}  = \int d^3r\,  
\sum_n \sum_k \frac{c_n(\als(M),\mu)}{M^{n}}  
\nn\\
&& \quad \times V_{n,k}(r,\mu^\prime, \mu) \; r^{k}  \times O_k(\mu^\prime,Mv^2,...) ,
\eea
where $O_k$ are the operators of pNRQCD that live at the low-energy scale
$Mv^2$, $\mu^\prime$ is the pNRQCD factorization scale and $V_{n,k}$
are the Wilson coefficients of the EFT that encode the contributions 
from the scale $r$ and are non-analytic in $r$.
Looking at the equations of motion of pNRQCD, we may identify 
$V_{n,0}$ with the $1/M^n$ potentials that enter the Schr\"odinger equation and 
$V_{n,k\neq 0}$ with the couplings of the ultrasoft degrees of freedom, which
provide corrections to the Schr\"odinger equation.

\subsection{The static QCD spectrum without light quarks}
The spectrum of a static quark and a static antiquark has been studied in QCD without light quarks 
on the lattice, for instance in \cite{Juge:2002br}. At short distances, it is well described by the
Coulomb potential in the colour-singlet or in the colour-octet configurations: 
$V_s = -4\als/3r$ and $V_o = \als/6r$. At large distances, the energies rise linearly with $r$.
Higher excitations develop a mass gap of order $\lQ$ with respect to the lowest one.
Reintroducing the heavy-quark mass $M$, the spectrum of the $Mv^2$ fluctuations around 
the lowest state constitutes the  quarkonium  spectrum while the spectrum of
the $Mv^2$ fluctuations around the higher excitations constitutes the hybrid spectrum.

\subsection{Quarkonia}
The energy of the lowest excitation between a static quark and a static antiquark is the quarkonium static energy.
Quarkonia may be identified with the $Mv^2$ spectrum that differentiates once $1/M$ corrections 
(first of all the kinetic energy) are added to the effective Lagrangian, which reads 
\bea
{\cal L}_{\rm pNRQCD}  &=& 
\int d^3r \, S^\dagger \left(i\partial_0 - \frac{p^2}{M} - V_s + ...\right) S 
\nn\\
&& 
+ \hbox{ultrasoft d.o.f.},
\label{pnrqcd}
\eea
where $S$ is the colour-singlet quarkonium field.

At short distances, the static potential is well described by perturbation theory 
up to NNNLL accuracy (for a recent analysis see \cite{Brambilla:2009bi}).
Therefore the lowest-lying quarkonium states have a radius that is small 
enough for perturbation theory to apply (see Fig.~\ref{figalphas}).
Higher-order corrections to the spectrum, masses and wave functions have been 
calculated in \cite{Kniehl:1999ud}.
Non-perturbative corrections are small and encoded in (local or non-local) condensates.
Many parameters and observables of the lowest quarkonium states have been calculated. They include:
$c$ and $b$ masses at NNLO and partially at N$^3$LO 
(for a review see \cite{Brambilla:2004wf}, a more recent result is in \cite{Pineda:2006gx});
$B_c$ mass at NNLO \cite{Brambilla:2000db}; 
$B_c^*$, $\eta_c$, $\eta_b$ masses at NLL \cite{Kniehl:2003ap}; 
quarkonium 1$P$ fine splittings at NLO \cite{Brambilla:2004wu}; 
$\Upsilon(1S)$, $\eta_b$ electromagnetic decays at NNLL \cite{Penin:2004ay};
$\Upsilon(1S)$ and $J/\psi$ radiative decays at NLO \cite{Fleming:2002sr};
$\Upsilon(1S) \to \gamma \eta_b$, $J/\psi \to \gamma \eta_c$ at NNLO \cite{Brambilla:2005zw};
$t\bar{t}$ cross section at NNLO \cite{Hoang:2001mm}; 
leading thermal effects on a quarkonium in medium: masses, widths,
\cite{Laine:2006ns}, ... .
For some recent reviews, we refer to \cite{Vairo:2006pc,Vairo:2006nq}.

High-lying quarkonia are the $Mv^2$ fluctuations around the long-range tail of the potential.
The long-range tail of the potential is not accessible by perturbation theory (see Fig.~\ref{figalphas}).
However, the potential may still be expanded in $1/M$ and each term of
the expansion can be expressed in terms of field-strength
insertions on a static Wilson loop \cite{Eichten:1980mw}, which can be
calculated on the lattice \cite{Koma:2006si}. The resulting potential may be
used in (\ref{pnrqcd}). The solution of the corresponding Schr\"odinger
equation provides the quarkonium masses and wave functions.
A tri\-vial example of application of this method is the mass of the $h_c$.
The lattice data show a vanishing long-range component 
of the spin-spin potential so that the potential appears to be entirely
dominated by its short-range, delta like, part. 
This suggests that the $^1P_1$ state should be close to 
the centre of gravity of the $^3P_J$ system.
Indeed, the measured mass of the $h_c$ by CLEO,  
$M_{h_c} = 3524.4 \pm 0.6 \pm 0.4~{\rm MeV}$ \cite{Rosner:2005ry}, 
and E835, $M_{h_c} = 3525.8 \pm 0.2 \pm 0.2~{\rm MeV}$ \cite{Andreotti:2005vu}, 
is just on the top of the mass of the $1^3P_J$ centre of gravity: 
$M_{\rm c.o.g.}(1^3P_J) = 3525.36 \pm 0.2 \pm 0.2~{\rm MeV}$.

\subsection{Gluonic excitations of quarkonia}
Many states, built on each of the hybrid potentials, are expected. 
These states typically develop a width also without including light quarks, 
since they may decay into lower states, e.g. like hybrid $\to$ glueball + quark-antiquark.

One possible candidate for such a state is the $Y(4260)$.
The $Y(4260)$ has been discovered by BABAR in the radiative return process 
$e^+e^- \to \gamma \pi^+\pi^- J/\psi$ with mass $M_Y = 4259 \pm 8^{+2}_{-6}~{\rm MeV}$
and width $\Gamma =  88 \pm 23 ^{+6}_{-4}~{\rm MeV}$  \cite{Aubert:2005rm}, 
and seen in the same process by BELLE
with mass $M_Y = 4247 \pm 12^{+17}_{-32}~{\rm MeV}$ and width 
$\Gamma =  108 \pm 19 \pm 10~{\rm MeV}$ \cite{belle:2007sj}
and by CLEO with mass $M_Y = 4284^{+17}_{-16}\pm 4~{\rm MeV}$ and width $\Gamma =  73^{+39}_{-25} \pm 5~{\rm MeV}$
\cite{He:2006kg}. CLEO has also confirmed the existence of an enhancement in the 
$\pi^+\pi^-J/\psi$ cross section at 4260 MeV in a measurement of direct $e^+e^-$ annihilation 
at $\sqrt{s}=4040$, $4160$ and $4260$ MeV \cite{Coan:2006rv}. 
The $Y(4260)$ $J^{PC}$ quantum numbers are $1^{--}$.
BABAR measures ${\cal B}(Y \to D \bar{D})/{\cal B}(Y \to J/\psi \pi^+\pi^-) < 1.0$
($\approx 500$ for $\psi(3770)$, which suggests an exotic interpretation 
for the $Y(4260)$) \cite{Aubert:2008pa}, moreover  ${\cal B}(Y \to D^* \bar{D})/{\cal B}(Y \to J/\psi
    \pi^+\pi^-) < 34$ and  ${\cal B}(Y \to D^* \bar{D}^*)/{\cal B}(Y \to
    J/\psi \pi^+\pi^-) < 40$ \cite{babar:2009xs}.

\begin{figure}[h]
\makebox[1.5truecm]{\phantom b}
\put(5,-90){\epsfxsize=5.5truecm \epsfbox{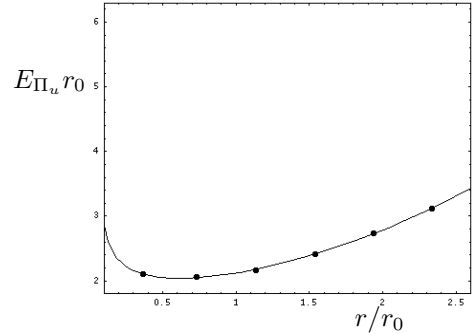}}
\put(108,-95){$r/r_0$}
\put(-20,-5){$E_{\Pi_u} r_0$}
\caption{The hybrid static potential $\Pi_u$ at short and intermediate
  distances, $r_0 \approx 0.5$ fm. The points are the lattice data
  from  \cite{Juge:2002br}, the continuous line is the fitting curve
$E_{\Pi_u}r_0 = \hbox{constant} + 0.11\, r_0/r  + 0.24\, (r/r_0)^2$.}
\label{figPiu}
\end{figure} 

Many interpretations have been proposed for the $Y(4260)$, one of this
is that the $Y(4260)$ is a charmonium hybrid \cite{Zhu:2005hp,Kou:2005gt,Close:2005iz}.
If the $Y(4260)$ is interpreted as a charmonium hybrid, one may rely on the heavy-quark expansion 
and on lattice calculations to study its properties. Decays into $D^{(*)} \bar{D}^{(*)}$ should be suppressed, since 
they are forbidden at leading order in the heavy-quark expansion \cite{Kou:2005gt}.
This is in agreement with the upper limit on $Y \to D \bar{D}$ reported by BABAR.
The quantum numbers of the  $Y(4260)$ are consistent with those of a
pseudoscalar $0^{-+}$ fluctuation $ |\phi\rangle$ belonging to the family of $Mv^2$ 
fluctuations around the gluonic excitation between a static quark and a static antiquark
with quantum numbers $1^{+-}$, also known as $\Pi_u$,  
\be
|Y\rangle  = |\Pi_u\rangle  \otimes  |\phi\rangle.
\ee
It is suggestive that, according to lattice calculations \cite{Juge:2002br}, 
$\Pi_u$ is the lowest gluonic excitation between a static quark and a static 
antiquark above the quark-antiquark colour singlet. $|\phi\rangle$  may be obtained 
as the  solution of the Schr\"odinger equation whose potential is the static 
energy of $\Pi_u$.
Fitting the static energy of $\Pi_u$ at short and intermediate distances, one gets 
$E_{\Pi_u}r_0 = \hbox{constant} + 0.11\, r_0/r  + 0.24\, (r/r_0)^2$, see Fig.~\ref{figPiu}.
Solving the corresponding Schr\"odinger equation, one gets 
$M_Y  = 2\times 1.48 + 0.87 + 0.53 = 4.36 ~~~\hbox{GeV}$, 
where 1.48 GeV is the charm mass in the RS scheme \cite{Pineda:2001zq} 
and 0.87 GeV is the gluelump mass in the same scheme \cite{Bali:2003jq}.

\subsection{The QCD spectrum with light quarks}
Adding light quarks changes the heavy quark-antiquark spectrum in the
following way  (see J.~Soto in \cite{Brambilla:2008zz}).
\begin{itemize}
\item[(1)]{We still have states made just of heavy quarks 
and gluons. They may develop a width because of the decay through pion
emission. If new states  made with heavy and 
light quarks develop a mass gap of order $\lQ$ with respect to 
the former ones, then these new states may be absorbed into the 
definition of the potentials or of the (local or non-local) condensates \cite{Brambilla:2002nu}.
}
\item[(2)]{
In addition, new states built using the light quark quantum numbers may form.
Possible states made of two heavy and light quarks include 
states built on the pair of heavy-light mesons ($D\bar{D}$, $B\bar{B}$, ...), 
molecular states \cite{Tornqvist:1991ks}, molecular states made of   
the usual quarkonium states, built on the static potential, and 
light hadrons (hadro-quarkonium) \cite{Dubynskiy:2008mq}, pairs of heavy-light
baryons \cite{Qiao:2005av}, tetraquark states \cite{Jaffe:1976ig} and likely many other states.
}
\end{itemize}

Clear evidence for four-quark states may be provided by a charged resonance, like the 
$Z^+(4430)$, $Z^+_1(4050)$ and $Z^+_2(4250)$ signals, detected by BELLE (but not confirmed 
by BABAR), possibly are. See \cite{Chistov09,Patrignani09} and the panel discussion 
at this conference. 

\begin{figure}[h]
\makebox[2cm]{\phantom b}
\put(20,0){{\epsfxsize=3truecm\epsffile{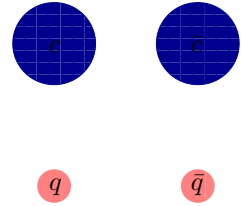}}}
\put(34,57){$c$}
\put(87,57){$\bar{c}$}
\put(34,5){$q$}
\put(87,5){$\bar{q}$}
\caption{$c\bar{c}q\bar{q}$ 4-quark state.}
\label{figQQqq1}
\end{figure}

There is accumulating evidence, although not yet conclusive,  
that the $X(3872)$ may be a four quark state.
The state $X(3872)$ has been discovered by BELLE in $B^\pm\to K^\pm X \to K^\pm \pi^+\pi^- J/\psi$ 
with $M_X = 3872.0 \pm 0.6 \pm 0.5~{\rm MeV}$ \cite{Choi:2003ue}, 
and confirmed by BABAR \cite{Aubert:2004ns} that measures  
$M_X = 3871.4 \pm 0.6 \pm 0.1 ~{\rm MeV}$ in $B^+\to K^+\pi^+\pi^- J/\psi$ 
and $M_X = 3868.7 \pm 1.5 \pm 0.4 ~{\rm MeV}$ in $B^0\to K^0\pi^+\pi^- J/\psi$ \cite{Aubert:2008gu}.
The state has also been seen at the Tevatron 
in $p\bar{p}\to X \to \pi^+\pi^- J/\psi$ by CDF with a mass $M_X = 3871.3 \pm
0.7 \pm 0.4 ~{\rm MeV}$ \cite{Acosta:2003zx} 
and by D0 with a mass $M_X = 3871.8 \pm 3.1 \pm 3.0~{\rm MeV}$ \cite{Abazov:2004kp}. 
BELLE has an upper limit on the width: $\Gamma < 2.3$ MeV, while BABAR finds 
 $\Gamma = 3.0^{+1.9}_{-1.4}\pm 0.9$ MeV \cite{Aubert:2007rva}. 
The $X(3872)$ has been detected in different decay modes, the decay 
$X \to D^0 \bar{D}^0 \pi^0$ being likely the dominant one:
${\cal B}(X \to D^0 \bar{D}^0 \pi^0)/{\cal B}(X\to  \pi^+\pi^-J/\psi) = 9.4^{+3.6}_{-4.3}$ 
\cite{Gokhroo:2006bt}. One should notice that 
BELLE finds a threshold enhancement peak in the $D^0 \bar{D}^0 \pi^0$ invariant mass 
at $3875.4 \pm 0.7^{+1.2}_{-2.0}$ MeV, which is about $2\,\sigma$ larger than the world-average 
mass of the $X(3872)$. The decay mode $X\to \gamma J/\psi$ \cite{Abe:2005ix} implies that the 
$X(3872)$ has positive charge conjugation. Ana\-lyses of angular distributions 
performed by BELLE \cite{Abe:2005iy} and CDF \cite{Kravchenko:2006qx}  favor 
a spin parity assignment $1^+$. The ratio ${\cal B}(X \to \pi^+\pi^-\pi^0 J/\psi)/{\cal B}(X\to
  \pi^+\pi^-J/\psi) = 1.0 \pm 0.4 \pm 0.3$ measured by BELLE \cite{Abe:2005ix}
suggests that the  $X(3872)$ is a mixture of isospin $I=1$ and $I=0$ states.
The substantial $I=1$ component requires that the $X(3872)$ contains 
$u\bar{u}$/$d\bar{d}$ pairs in addition to hidden charm, which thus 
qualifies it as a four-quark state \cite{Voloshin:2006wf}.
Hence, most recently, the majority of theoretical studies has  
analyzed the $X(3872)$ as a four-quark state with $J^{PC}$ quantum numbers $1^{++}$.
See Fig.~\ref{figQQqq1}.

Three quark-pair configurations are possible for a four-quark state of the type 
$c\bar{c}q\bar{q}$ ($q$ stands for a generic light quark).
All of them have been exploited in the literature. However, the resulting models are not equivalent, 
because different dynamics are attributed to different confi\-gurations.

Clearly, for states made of heavy quark-antiquark pairs and light quarks, it would be 
useful to have the spectrum of tetraquark potentials, like the one that we have 
for the gluonic excitations and that we discussed above. It would allow 
us to build a plethora of states on each of the potentials, many of them developing a 
width due to decays through pion (or other light hadron) emission. 
Diquarks have been recently investigated on the lattice \cite{Alexandrou:2006cq}.
In the lack of further theoretical input from QCD, many tetraquark 
studies rely on phenomenological models of the tetraquark interaction unless 
some special hierarchy of dynamical scales may be further exploited on the 
top of the non-relativistic and perturbative expansions discussed so far.

\begin{figure}[h]
\makebox[2cm]{\phantom b}
\put(20,0){{\epsfxsize=3truecm\epsffile{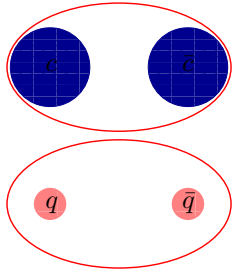}}}
\put(35,75){$c$}
\put(86,75){$\bar{c}$}
\put(35,23){$q$}
\put(86,23){$\bar{q}$}
\caption{$c\bar{c}q\bar{q}$ 4-quark state in the configuration of \cite{Hogaasen:2005jv}.}
\label{figQQqq3}
\end{figure}

In \cite{Hogaasen:2005jv}, see Fig.~\ref{figQQqq3}, 
it is assumed that $X \sim (c\bar{c})^8_{S=1}   (q\bar{q})^8_{S=1}$, 
i.e. that the dominant Fock-space component contains a $c\bar{c}$ pair and a $q\bar{q}$ pair
in a colour-octet configuration with spin 1. 
Calculations have been based on a phenomenological interaction Hamiltonian. 

\begin{figure}[h]
\makebox[2cm]{\phantom b}
\put(20,0){{\epsfxsize=3truecm\epsffile{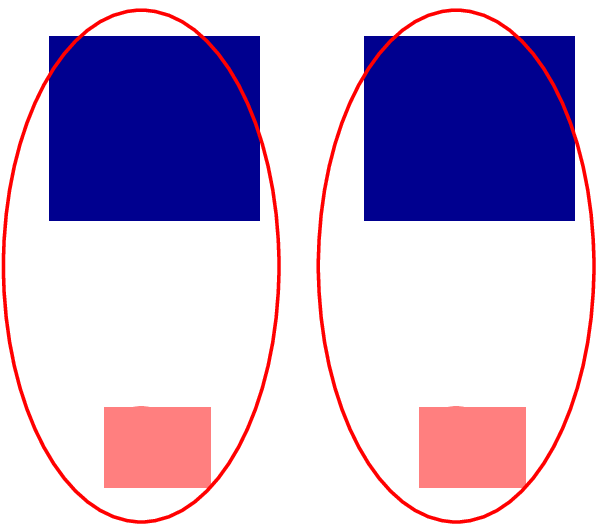}}}
\put(38,55){$c$}
\put(83,55){$\bar{c}$}
\put(38,10){$q$}
\put(83,10){$\bar{q}$}
\caption{$c\bar{c}q\bar{q}$ 4-quark state in the configuration of \cite{Maiani:2004vq}.}
\label{figQQqq2}
\end{figure}

In \cite{Maiani:2004vq}, see Fig.~\ref{figQQqq2}, 
it is assumed that $X \sim (cq)^{\bar{3}}_{S=1}   (\bar{c}\bar{q})^{3}_{S=0}
+ (cq)^{\bar{3}}_{S=0}   (\bar{c}\bar{q})^{3}_{S=1}$. 
Here, the clustering of quark pairs in tightly bound colour-triplet diquarks 
is not induced by a scale separation as it would happen in baryons made of two 
heavy quarks \cite{Brambilla:2005yk},  but is a dynamical assumption of the model. 
In particular, the model predicts the existence of two neutral 
states made of $cu\bar{c}\bar{u}$ ($X_u$) and $cd\bar{c}\bar{d}$ ($X_d$)
and of two charged ones. 
The two resonances discovered by BELLE and BABAR, the first decaying in $J/\psi\, \pi^+\pi^-$ 
and the second preferably in $D^0\bar D^0 \pi^0$ have been suggested 
as possible candidates for the $X_d$ and $X_u$ \cite{Maiani:2007vr}.

\begin{figure}[h]
\makebox[2cm]{\phantom b}
\put(20,0){{\epsfxsize=3truecm\epsffile{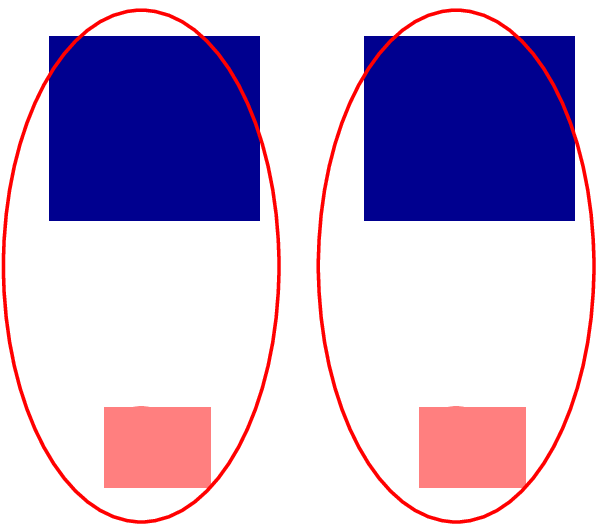}}}
\put(38,55){$c$}
\put(83,55){$\bar{c}$}
\put(38,10){$\bar{q}$}
\put(83,10){$q$}
\caption{$c\bar{c}q\bar{q}$ 4-quark state in the configuration of \cite{Tornqvist:1993ng,Swanson:2003tb}.}
\label{figQQqq2bis}
\end{figure}

In \cite{Tornqvist:1993ng,Swanson:2003tb}, see Fig.~\ref{figQQqq2bis}, it is assumed that
$X \sim (c\bar{q})^1_{S=0}   (q\bar{c})^1_{S=1} 
+ (c\bar{q})^1_{S=1}   (q\bar{c})^1_{S=0}$ $ \sim D^0\,\bar{D}^{*\,0} + D^{*\,0}\, \bar{D}^0$, 
i.e. that the dominant Fock-space component of the $X(3872)$ is a $D^0\, \bar{D}^{*\,0}$ and 
$D^{*\,0} \, {\bar D}^0$ molecule; small short-range components 
of the type $(c\bar{c})^1_{S=1}   (q\bar{q})^1_{S=1}$ $\sim$ $J/\psi \,\rho, \omega$ are
included as well. Predictions depend on the adopted phenomenological
Hamiltonian, which typically contains, in the short range ($\sim 1/\lQ$), 
potential-type interactions among the quarks and,  
in the long range ($\sim 1/m_\pi$), the one-pion exchange. 
The prediction $\Gamma(X\to \pi^+\pi^-J/\psi) \approx \Gamma(X\to
  \pi^+\pi^-\pi^0 J/\psi)$ made in \cite{Swanson:2003tb} turned out to be
  consistent with the BELLE result \cite{Abe:2005ix}. 
However, another prediction, $\Gamma(X\to \pi^+\pi^-J/\psi) \approx 20 \,\Gamma(X\to
  D^0 \bar{D}^0 \pi^0)$, is two orders of magnitude far from 
the data  \cite{Gokhroo:2006bt}. Not necessarily this points to a failure of
the molecular model, but possibly to a smaller $J/\psi \,\rho$ component in the $X(3872)$ Fock space.

\begin{figure}[h]
\makebox[1.5cm]{\phantom b}
\put(20,0){{\epsfxsize=4.6truecm\epsffile{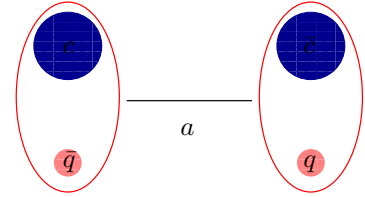}}}
\put(38,53){$c$}
\put(128,53){$\bar{c}$}
\put(38,10){$\bar{q}$}
\put(128,10){$q$}
\put(82,22){$a$}
\caption{$c\bar{c}q\bar{q}$ 4-quark state in the configuration of 
\cite{Pakvasa:2003ea,Voloshin:2003nt,Braaten:2003he}.}
\label{figQQqq4}
\end{figure}

In \cite{Pakvasa:2003ea,Voloshin:2003nt,Braaten:2003he}, see Fig.~\ref{figQQqq4}, 
it is assumed not only that the $X(3872)$ is a $D^0 \, \bar{D}^{*\,0}$ and  
${\bar D}^0 \, D^{*\,0}$ molecule, but also that it is loosely bound, 
i.e. that the following hierarchy of scales is realized:
$\lQ \gg m_\pi$ $\gg m_\pi^2/M_{D_0}$ $\approx$  $10 ~{\rm MeV}$ $\gg E_{\rm binding}$. 
Indeed, the binding energy, $E_{\rm binding}$, which may be estimated from $M_X - (M_{D^{*\,0}}+M_{D^{0}})$,
is very close to zero, i.e. much smaller than the natural scale 
$m_\pi^2/M_{D_0}$. This is also the case when using a recent CLEO determination 
of the $D_0$ mass, $M_{D_0} = 1864.847\pm 0.150\pm 0.095$ MeV \cite{Cawlfield:2007dw}. 
The main uncertainty comes from the 
$X(3872)$ mass. Systems with a short-range interaction 
and a large scattering length have universal properties that may be exploited: 
in particular, production and decay amplitudes factorize in a short-range 
and a long-range part, where the latter depends only on one single parameter, 
the scattering length.  
The long-range molecular $D^0 \, \bar{D}^{*\,0}$ and  ${\bar D}^0 \, D^{*\,0}$ components  
of the $X(3872)$ should be responsible for the $X(3872)$ decaying into 
$D^0 \bar{D}^0 \pi^0$. For a recent analysis of the  BELLE data about 
the $D^0 \bar{D}^0 \pi^0$ final state enhancement and the molecular picture we
refer to \cite{Hanhart:2007yq}.
For discussion about the evaluation of the $X(3872)$ production cross section 
at the Tevatron inside the molecular model we refer to \cite{Bignamini:2009sk,Artoisenet:2009wk}.

\subsection{ Coupled channels}
An important (and yet unsolved) problem is how all the different kind of states 
(with and without light quarks) interact with each other. 
A systematic treatment does not exist so far. 
For the coupling with two-meson states, most of the existing analyses rely on
two models, which are now more than 30 years old:
the Cornell coupled-channel model \cite{Eichten:1978tg} and the $^3P_0$ model \cite{Le Yaouanc:1972ae}.
Steps towards a lattice based approach have been undertaken recently 
\cite{Bali:2005fu,Bali09} and may, in perspective, provide an alternative, QCD
based, treatment.

\section{Conclusions}
Our understanding of how a (effective field) theory of quarkonium should look like
has dramatically increased over the last decade.

For states below threshold such a theory exists and allows a systematic 
study of the quarkonium lowest resonances. Even precision physics is possible.
Higher resonances may need to be supplemented by lattice data.
High-quality lattice data have become available in the last years for some 
crucial quantities (e.g. potentials, decay matrix elements, ...).

For states above threshold, the picture appears much more uncertain. Many degrees of freedom 
seem to be present, and the absence of a clear systematics is an obstacle to 
an universal picture. Most likely, systematic descriptions  will be 
found that suite only specific families of states, the near-threshold
molecular states providing an example.


\begin{thebibliography}{99} 
%\cite{Caswell:1985ui}
\bibitem{Caswell:1985ui}
  W.~E.~Caswell and G.~P.~Lepage,
  %``Effective Lagrangians For Bound State Problems In QED, QCD, And Other Field Theories,''
  Phys.\ Lett.\ B {\bf 167} (1986) 437;
  %%CITATION = PHLTA,B167,437;%%
%\cite{Bodwin:1994jh}
%\bibitem{Bodwin:1994jh}
  G.~T.~Bodwin, E.~Braaten and G.~P.~Lepage,
  %``Rigorous QCD analysis of inclusive annihilation and production of heavy
  %quarkonium,''
  Phys.\ Rev.\ D {\bf 51} (1995) 1125 
  [Erratum-ibid.\ D {\bf 55} (1997) 5853]
  [hep-ph/9407339].
  %%CITATION = HEP-PH 9407339;%%

%\cite{Pineda:1997bj}
\bibitem{Pineda:1997bj}
  A.~Pineda and J.~Soto,
  %``Effective field theory for ultrasoft momenta in NRQCD and NRQED,''
  Nucl.\ Phys.\ Proc.\ Suppl.\  {\bf 64} (1998) 428 
  [arXiv:hep-ph/9707481];
  %%CITATION = HEP-PH 9707481;%%
%\cite{Brambilla:1999xf}
%\bibitem{Brambilla:1999xf}
  N.~Brambilla,  A.~Pineda, J.~Soto and A.~Vairo,
  %``Potential NRQCD: An effective theory for heavy quarkonium,''
  Nucl.\ Phys.\ B {\bf 566} (2000) 275 [arXiv:hep-ph/9907240].
  %%CITATION = HEP-PH 9907240;%%

%\cite{Neubert:1993mb}
\bibitem{Neubert:1993mb}
  M.~Neubert,
  %``Heavy quark symmetry,''
  Phys.\ Rept.\  {\bf 245} (1994) 259 
  [arXiv:hep-ph/9306320];
  %%CITATION = HEP-PH 9306320;%%
%\cite{Manohar:2000dt}
%\bibitem{Manohar:2000dt}
  A.~V.~Manohar and M.~B.~Wise,
  %``Heavy quark physics,''
  Camb.\ Monogr.\ Part.\ Phys.\ Nucl.\ Phys.\ Cosmol.\  {\bf 10} (2000) 1. 
  %%CITATION = CMPCE,10,1;%%

%\cite{Brambilla:2004jw}
\bibitem{Brambilla:2004jw}
  N.~Brambilla, A.~Pineda, J.~Soto and A.~Vairo,
  %``Effective field theories for heavy quarkonium,''
  Rev.\ Mod.\ Phys.\  {\bf 77} (2005) 1423
  [arXiv:hep-ph/0410047].
  %%CITATION = RMPHA,77,1423;%%

%\cite{Vairo:2009rs}
\bibitem{Vairo:2009rs}
  A.~Vairo,
  %``Non-relativistic bound states: the long way back from the Bethe-Salpeter to
  %the Schroedinger equation,''
  arXiv:0902.3346 [hep-ph].
  %%CITATION = ARXIV:0902.3346;%%

%\cite{Godfrey:1985xj}
\bibitem{Godfrey:1985xj}
  S.~Godfrey and N.~Isgur,
  %``Mesons In A Relativized Quark Model With Chromodynamics,''
  Phys.\ Rev.\  D {\bf 32} (1985) 189.
  %%CITATION = PHRVA,D32,189;%%

%\cite{Matsui:1986dk}
\bibitem{Matsui:1986dk}
  T.~Matsui and H.~Satz,
  %``J/psi Suppression by Quark-Gluon Plasma Formation,''
  Phys.\ Lett.\  B {\bf 178} (1986) 416.
  %%CITATION = PHLTA,B178,416;%%

%\cite{Petreczky09}
\bibitem{Petreczky09}
  P.~Petreczky, at this conference.

%\cite{Brambilla:2006ph}
\bibitem{Brambilla:2006ph}
  N.~Brambilla, E.~Mereghetti and A.~Vairo,
  %``Electromagnetic quarkonium decays at order v**7,''
  JHEP {\bf 0608} (2006) 039
  [arXiv:hep-ph/0604190].
  %%CITATION = JHEPA,0608,039;%%

%\cite{Brambilla:2008zg}
\bibitem{Brambilla:2008zg}
  N.~Brambilla, E.~Mereghetti and A.~Vairo,
  %``Hadronic quarkonium decays at order v^7,''
  Phys.\ Rev.\  D {\bf 79} (2009) 074002
  [arXiv:0810.2259 [hep-ph]].
  %%CITATION = PHRVA,D79,074002;%%

%\cite{Bodwin:2005gg}
\bibitem{Bodwin:2005gg}
  G.~T.~Bodwin, J.~Lee and D.~K.~Sinclair,
  %``Spin correlations and velocity-scaling in color-octet NRQCD matrix
  %elements,''
  Phys.\ Rev.\  D {\bf 72} (2005) 014009
  [arXiv:hep-lat/0503032].
  %%CITATION = PHRVA,D72,014009;%%

%\cite{Brambilla:2004wf}
\bibitem{Brambilla:2004wf}
  N.~Brambilla {\it et al.},
  Heavy quarkonium physics,
  CERN-2005-005, (CERN, Geneva, 2005)
  [arXiv:hep-ph/0412158].
  %%CITATION = HEP-PH 0412158;%%

%\cite{Amsler:2008zzb}
\bibitem{Amsler:2008zzb}
  C.~Amsler {\it et al.}  [Particle Data Group],
  %``Review of particle physics,''
  Phys.\ Lett.\  B {\bf 667} (2008) 1.
  %%CITATION = PHLTA,B667,1;%%

%\cite{Brambilla:2007cz}
\bibitem{Brambilla:2007cz}
  N.~Brambilla, X.~Garcia i Tormo, J.~Soto and A.~Vairo,
  %``Extraction of alpha_s from radiative Upsilon(1S) decays,''
  Phys.\ Rev.\  D {\bf 75} (2007) 074014
  [arXiv:hep-ph/0702079].
  %%CITATION = PHRVA,D75,074014;%%

%\cite{Artoisenet09}
\bibitem{Artoisenet09}
  P.~Artoisenet, at this conference.

%\cite{Juge:2002br}
\bibitem{Juge:2002br}
  K.~J.~Juge, J.~Kuti and C.~Morningstar,
  %``Fine structure of the QCD string spectrum,''
  Phys.\ Rev.\ Lett.\  {\bf 90} (2003) 161601
  [arXiv:hep-lat/0207004].
  %%CITATION = PRLTA,90,161601;%%

%\cite{Brambilla:2009bi}
\bibitem{Brambilla:2009bi}
  N.~Brambilla, A.~Vairo, X.~Garcia i Tormo and J.~Soto,
  %``The QCD static energy at NNNLL,''
  Phys.\ Rev.\  D {\bf 80} (2009) 034016
  [arXiv:0906.1390 [hep-ph]].
  %%CITATION = PHRVA,D80,034016;%%

%\cite{Kniehl:1999ud}
\bibitem{Kniehl:1999ud}
  B.~A.~Kniehl and A.~A.~Penin,
  %``Ultrasoft effects in heavy quarkonium physics,''
  Nucl.\ Phys.\  B {\bf 563} (1999) 200
  [arXiv:hep-ph/9907489];
  %%CITATION = NUPHA,B563,200;%%
%\cite{Brambilla:1999xj}
%\bibitem{Brambilla:1999xj}
  N.~Brambilla, A.~Pineda, J.~Soto and A.~Vairo,
  %``The heavy quarkonium spectrum at order m alpha(s)**5 ln(alpha(s)),''
  Phys.\ Lett.\  B {\bf 470} (1999) 215
  [arXiv:hep-ph/9910238];
  %%CITATION = PHLTA,B470,215;%%
%\cite{Kniehl:1999mx}
%\bibitem{Kniehl:1999mx}
  B.~A.~Kniehl and A.~A.~Penin,
  %``Order alpha(s)**3 ln**2(1/alpha(s)) corrections to heavy-quarkonium
  %creation and annihilation,''
  Nucl.\ Phys.\  B {\bf 577} (2000) 197
  [arXiv:hep-ph/9911414];
  %%CITATION = NUPHA,B577,197;%%
%\cite{Kniehl:2002br}
%\bibitem{Kniehl:2002br}
  B.~A.~Kniehl, A.~A.~Penin, V.~A.~Smirnov and M.~Steinhauser,
  %``Potential NRQCD and heavy-quarkonium spectrum at
  %next-to-next-to-next-to-leading order,''
  Nucl.\ Phys.\  B {\bf 635} (2002) 357
  [arXiv:hep-ph/0203166];
  %%CITATION = NUPHA,B635,357;%%
%\cite{Penin:2002zv}
%\bibitem{Penin:2002zv}
  A.~A.~Penin and M.~Steinhauser,
  %``Heavy Quarkonium Spectrum at ${\cal O}(\alpha_s^5m_q)$ and Bottom/Top Quark
  %Mass Determination,''
  Phys.\ Lett.\  B {\bf 538} (2002) 335
  [arXiv:hep-ph/0204290];
  %%CITATION = PHLTA,B538,335;%%
%\cite{Beneke:2007gj}
%\bibitem{Beneke:2007gj}
  M.~Beneke, Y.~Kiyo and K.~Schuller,
  %``Third-order non-Coulomb correction to the S-wave quarkonium wave
  %functions at the origin,''
  Phys.\ Lett.\  B {\bf 658} (2008) 222
  [arXiv:0705.4518 [hep-ph]];
  %%CITATION = PHLTA,B658,222;%%
%\cite{Beneke:2007pj}
%\bibitem{Beneke:2007pj}
  M.~Beneke, Y.~Kiyo and A.~A.~Penin,
  %``Ultrasoft contribution to quarkonium production and annihilation,''
  Phys.\ Lett.\  B {\bf 653} (2007) 53
  [arXiv:0706.2733 [hep-ph]].
  %%CITATION = PHLTA,B653,53;%%

%\cite{Pineda:2006gx}
\bibitem{Pineda:2006gx}
  A.~Pineda and A.~Signer,
  %``Renormalization Group Improved Sum Rule Analysis for the Bottom Quark
  %Mass,''
  Phys.\ Rev.\  D {\bf 73} (2006) 111501
  [arXiv:hep-ph/0601185].
  %%CITATION = PHRVA,D73,111501;%%

%\cite{Brambilla:2000db}
\bibitem{Brambilla:2000db}
  N.~Brambilla and A.~Vairo,
  %``The $B_c$ mass up to order $\alpha($s$)^{4}$,''
  Phys.\ Rev.\  D {\bf 62} (2000) 094019
  [arXiv:hep-ph/0002075];
  %%CITATION = PHRVA,D62,094019;%%
%\cite{Brambilla:2001fw}
%\bibitem{Brambilla:2001fw}
  N.~Brambilla, Y.~Sumino and A.~Vairo,
  %``Quarkonium spectroscopy and perturbative QCD: A new perspective,''
  Phys.\ Lett.\  B {\bf 513} (2001) 381
  [arXiv:hep-ph/0101305];
  %%CITATION = PHLTA,B513,381;%%
%\cite{Brambilla:2001qk}
%\bibitem{Brambilla:2001qk}
  N.~Brambilla, Y.~Sumino and A.~Vairo,
  %``Quarkonium spectroscopy and perturbative QCD: Massive quark loop
  %effects,''
  Phys.\ Rev.\  D {\bf 65} (2002) 034001
  [arXiv:hep-ph/0108084].
  %%CITATION = PHRVA,D65,034001;%%

%\cite{Kniehl:2003ap}
\bibitem{Kniehl:2003ap}
  B.~A.~Kniehl, A.~A.~Penin, A.~Pineda, V.~A.~Smirnov and M.~Steinhauser,
  %``$M(\eta_b)$ and $\alpha_s$ from Nonrelativistic Renormalization Group,''
  Phys.\ Rev.\ Lett.\  {\bf 92} (2004) 242001
  [arXiv:hep-ph/0312086];
  %%CITATION = PRLTA,92,242001;%%
%\cite{Penin:2004xi}
%\bibitem{Penin:2004xi}
  A.~A.~Penin, A.~Pineda, V.~A.~Smirnov and M.~Steinhauser,
  %``M(B*($c$) ) - M(B($c$) ) splitting from nonrelativistic renormalization
  %group,''
  Phys.\ Lett.\  B {\bf 593} (2004) 124
  [Erratum-ibid.\  {\bf 677} (2009) 343]
  [arXiv:hep-ph/0403080].
  %%CITATION = PHLTA,B593,124;%%

%\cite{Brambilla:2004wu}
\bibitem{Brambilla:2004wu}
  N.~Brambilla and A.~Vairo,
  %``The 1P quarkonium fine splittings at NLO,''
  Phys.\ Rev.\  D {\bf 71} (2005) 034020
  [arXiv:hep-ph/0411156].
  %%CITATION = PHRVA,D71,034020;%%

%\cite{Penin:2004ay}
\bibitem{Penin:2004ay}
  A.~A.~Penin, A.~Pineda, V.~A.~Smirnov and M.~Steinhauser,
  %``Spin dependence of heavy quarkonium production and annihilation rates:
  %Complete next-to-next-to-leading logarithmic result,''
  Nucl.\ Phys.\  B {\bf 699} (2004) 183
  [arXiv:hep-ph/0406175].
  %%CITATION = NUPHA,B699,183;%%

%\cite{Brambilla:2005zw}
\bibitem{Brambilla:2005zw}
  N.~Brambilla, Y.~Jia and A.~Vairo,
  %``Model-independent study of magnetic dipole transitions in quarkonium,''
  Phys.\ Rev.\  D {\bf 73} (2006) 054005
  [arXiv:hep-ph/0512369].
  %%CITATION = PHRVA,D73,054005;%%

%\cite{Fleming:2002sr}
\bibitem{Fleming:2002sr}
  S.~Fleming and A.~K.~Leibovich,
  %``The photon spectrum in Upsilon decays,''
  Phys.\ Rev.\  D {\bf 67} (2003) 074035
  [arXiv:hep-ph/0212094];
  %%CITATION = PHRVA,D67,074035;%%
%\cite{GarciaiTormo:2005ch}
%\bibitem{GarciaiTormo:2005ch}
  X.~Garcia i Tormo and J.~Soto,
  %``Semi-inclusive radiative decays of Upsilon(1S),''
  Phys.\ Rev.\  D {\bf 72} (2005) 054014
  [arXiv:hep-ph/0507107];
  %%CITATION = PHRVA,D72,054014;%%
%\cite{GarciaiTormo:2005bs}
%\bibitem{GarciaiTormo:2005bs}
  X.~Garcia i Tormo and J.~Soto,
  %``Radiative decays and the nature of heavy quarkonia,''
  Phys.\ Rev.\ Lett.\  {\bf 96} (2006) 111801
  [arXiv:hep-ph/0511167].
  %%CITATION = PRLTA,96,111801;%%

%\cite{Hoang:2001mm}
\bibitem{Hoang:2001mm}
  A.~H.~Hoang, A.~V.~Manohar, I.~W.~Stewart and T.~Teubner,
  %``The threshold t anti-t cross section at NNLL order,''
  Phys.\ Rev.\  D {\bf 65} (2002) 014014
  [arXiv:hep-ph/0107144];
  %%CITATION = PHRVA,D65,014014;%%
%\cite{Hoang:2003xg}
%\bibitem{Hoang:2003xg}
  A.~H.~Hoang,
  %``Top Pair Production at Threshold and Effective Theories,''
  Acta Phys.\ Polon.\  B {\bf 34} (2003) 4491
  [arXiv:hep-ph/0310301];
  %%CITATION = APPOA,B34,4491;%%
%\cite{Hoang:2006pc}
%\bibitem{Hoang:2006pc}
  A.~H.~Hoang,
  %``Top threshold physics,''
  PoS {\bf TOP2006} (2006) 032
  [arXiv:hep-ph/0604185];
  %%CITATION = POSCI,TOP2006,032;%%
%\cite{Pineda:2006ri}
%\bibitem{Pineda:2006ri}
  A.~Pineda and A.~Signer,
  %``Heavy quark pair production near threshold with potential  non-relativistic
  %QCD,''
  Nucl.\ Phys.\  B {\bf 762} (2007) 67
  [arXiv:hep-ph/0607239].
  %%CITATION = NUPHA,B762,67;%%

%\cite{Laine:2006ns}
\bibitem{Laine:2006ns}
  M.~Laine, O.~Philipsen, P.~Romatschke and M.~Tassler,
  %``Real-time static potential in hot QCD,''
  JHEP {\bf 0703} (2007) 054
  [arXiv:hep-ph/0611300];
  %%CITATION = JHEPA,0703,054;%%
%\cite{Beraudo:2007ky}
%\bibitem{Beraudo:2007ky}
  A.~Beraudo, J.~P.~Blaizot and C.~Ratti,
  %``Real and imaginary-time $Q\bar{Q}$ correlators in a thermal medium,''
  Nucl.\ Phys.\  A {\bf 806} (2008) 312
  [arXiv:0712.4394 [nucl-th]];
  %%CITATION = NUPHA,A806,312;%%
%\cite{Brambilla:2008cx}
%\bibitem{Brambilla:2008cx}
  N.~Brambilla, J.~Ghiglieri, A.~Vairo and P.~Petreczky,
  %``Static quark-antiquark pairs at finite temperature,''
  Phys.\ Rev.\  D {\bf 78} (2008) 014017
  [arXiv:0804.0993 [hep-ph]].
  %%CITATION = PHRVA,D78,014017;%%

%\cite{Vairo:2006pc}
\bibitem{Vairo:2006pc}
  A.~Vairo,
  %``Heavy hadron spectroscopy,''
  Int.\ J.\ Mod.\ Phys.\  A {\bf 22} (2007) 5481
  [arXiv:hep-ph/0611310].
  %%CITATION = IMPAE,A22,5481;%%

%\cite{Vairo:2006nq}
\bibitem{Vairo:2006nq}
  A.~Vairo,
  %``Heavy quarkonium physics from effective field theories,''
  Eur.\ Phys.\ J.\  A {\bf 31} (2007) 728
  [arXiv:hep-ph/0610251].
  %%CITATION = EPHJA,A31,728;%%

%\cite{Eichten:1980mw}
\bibitem{Eichten:1980mw}
  E.~Eichten and F.~Feinberg,
  %``Spin Dependent Forces In QCD,''
  Phys.\ Rev.\  D {\bf 23} (1981) 2724;
  %%CITATION = PHRVA,D23,2724;%%
%\cite{Barchielli:1988zp}
%\bibitem{Barchielli:1988zp}
  A.~Barchielli, N.~Brambilla and G.~M.~Prosperi,
  %``Relativistic Corrections To The Quark - Anti-Quark Potential And The
  %Quarkonium Spectrum,''
  Nuovo Cim.\  A {\bf 103} (1990) 59;
  %%CITATION = NUCIA,A103,59;%%
%\cite{Brambilla:2000gk}
%\bibitem{Brambilla:2000gk}
  N.~Brambilla, A.~Pineda, J.~Soto and A.~Vairo,
  %``The QCD potential at O(1/m),''
  Phys.\ Rev.\  D {\bf 63} (2001) 014023
  [arXiv:hep-ph/0002250];
  %%CITATION = PHRVA,D63,014023;%%
%\cite{Pineda:2000sz}
%\bibitem{Pineda:2000sz}
  A.~Pineda and A.~Vairo,
  %``The QCD potential at O (1 / $m^{2)}$ : Complete spin dependent and spin
  %independent result,''
  Phys.\ Rev.\  D {\bf 63} (2001) 054007
  [Erratum-ibid.\  D {\bf 64} (2001) 039902]
  [arXiv:hep-ph/0009145];
  %%CITATION = PHRVA,D63,054007;%%
%\cite{Brambilla:2003mu}
%\bibitem{Brambilla:2003mu}
  N.~Brambilla, A.~Pineda, J.~Soto and A.~Vairo,
  %``The (m Lambda(QCD))**1/2 scale in heavy quarkonium,''
  Phys.\ Lett.\  B {\bf 580} (2004) 60
  [arXiv:hep-ph/0307159].
  %%CITATION = PHLTA,B580,60;%%

%\cite{Koma:2006si}
\bibitem{Koma:2006si}
  Y.~Koma, M.~Koma and H.~Wittig,
  %``Nonperturbative determination of the QCD potential at O(1/m),''
  Phys.\ Rev.\ Lett.\  {\bf 97} (2006) 122003
  [arXiv:hep-lat/0607009];
  %%CITATION = PRLTA,97,122003;%%
%\cite{Koma:2006fw}
%\bibitem{Koma:2006fw}
  Y.~Koma and M.~Koma,
  %``Spin-dependent potentials from lattice QCD,''
  Nucl.\ Phys.\  B {\bf 769} (2007) 79
  [arXiv:hep-lat/0609078];
  %%CITATION = NUPHA,B769,79;%%
%\cite{Koma:2007jq}
%\bibitem{Koma:2007jq}
  Y.~Koma, M.~Koma and H.~Wittig,
  %``Relativistic corrections to the static potential at O(1/m) and O(1/m^2),''
  PoS {\bf LAT2007} (2007) 111
  [arXiv:0711.2322 [hep-lat]];
  %%CITATION = POSCI,LAT2007,111;%%
%\cite{Koma:2009ws}
%\bibitem{Koma:2009ws}
  Y.~Koma and M.~Koma,
  %``Scaling study of the relativistic corrections to the static potential,''
  arXiv:0911.3204 [hep-lat].
  %%CITATION = ARXIV:0911.3204;%%

%\cite{Rosner:2005ry}
\bibitem{Rosner:2005ry}
  J.~L.~Rosner {\it et al.}  [CLEO Collaboration],
  %``Observation of h/c ((1)P(1)) state of charmonium,''
  Phys.\ Rev.\ Lett.\  {\bf 95} (2005) 102003
  [arXiv:hep-ex/0505073].
  %%CITATION = PRLTA,95,102003;%%

%\cite{Andreotti:2005vu}
\bibitem{Andreotti:2005vu}
  M.~Andreotti {\it et al.},
  %``Results of a search for the h(c) (1)P(1) state of charmonium in the eta(c)
  %gamma and J/psi pi0 decay modes,''
  Phys.\ Rev.\  D {\bf 72} (2005) 032001.
  %%CITATION = PHRVA,D72,032001;%%

%\cite{Aubert:2005rm}
\bibitem{Aubert:2005rm}
  B.~Aubert {\it et al.}  [BABAR Collaboration],
  % ``Observation of a broad structure in the $\pi^+ \pi^- J/\psi$ mass spectrum
  %around 4.26-GeV/c$^2$,''
  Phys.\ Rev.\ Lett.\  {\bf 95} (2005) 142001 
  [arXiv:hep-ex/0506081].
  %%CITATION = HEP-EX 0506081;%%

%\cite{belle:2007sj}
\bibitem{belle:2007sj}
  C.~Z.~Yuan {\it et al.}  [Belle Collaboration],
  %``Measurement of $e^+e^- \to \pi^+\pi^-J/\psi$ Cross Section via Initial
  %State Radiation at Belle,''
  Phys.\ Rev.\ Lett.\  {\bf 99} (2007) 182004
  [arXiv:0707.2541 [hep-ex]].
  %%CITATION = PRLTA,99,182004;%%

%\cite{He:2006kg}
\bibitem{He:2006kg}
  Q.~He {\it et al.}  [CLEO Collaboration],
  %``Confirmation of the Y(4260) resonance production in ISR,''
  Phys.\ Rev.\  D {\bf 74} (2006) 091104
  [arXiv:hep-ex/0611021].
  %%CITATION = PHRVA,D74,091104;%%

%\cite{Coan:2006rv}
\bibitem{Coan:2006rv}
  T.~E.~Coan {\it et al.}  [CLEO Collaboration],
  %``Charmonium decays of Y(4260), psi(4160), and psi(4040),''
  Phys.\ Rev.\ Lett.\  {\bf 96} (2006) 162003 
  [arXiv:hep-ex/0602034].
  %%CITATION = HEP-EX 0602034;%%

%\cite{Aubert:2008pa}
\bibitem{Aubert:2008pa}
  B.~Aubert {\it et al.}  [BABAR Collaboration],
  %``Study of the Exclusive Initial-State-Radiation Production of the DDbar
  %System,''
  arXiv:0710.1371 [hep-ex].
  %%CITATION = ARXIV:0710.1371;%%

%\cite{babar:2009xs}
\bibitem{babar:2009xs}
  B.~Aubert {\it et al.}  [BABAR Collaboration],
  %``Exclusive Initial-State-Radiation Production of the $D \bar D$, $D \bar
  %D^*$, and $D^* \bar D^*$, Systems,''
  Phys.\ Rev.\  D {\bf 79} (2009) 092001
  [arXiv:0903.1597 [hep-ex]].
  %%CITATION = PHRVA,D79,092001;%%

%\cite{Zhu:2005hp}
\bibitem{Zhu:2005hp}
  S.~L.~Zhu,
  %``The possible interpretations of Y(4260),''
  Phys.\ Lett.\ B {\bf 625} (2005) 212 
  [arXiv:hep-ph/0507025].
  %%CITATION = HEP-PH 0507025;%%

%\cite{Kou:2005gt}
\bibitem{Kou:2005gt}
  E.~Kou and O.~Pene,
  %``Suppressed decay into open charm for the Y(4260) being an hybrid,''
  Phys.\ Lett.\ B {\bf 631} (2005) 164 
  [arXiv:hep-ph/0507119].
  %%CITATION = HEP-PH 0507119;%%

%\cite{Close:2005iz}
\bibitem{Close:2005iz}
  F.~E.~Close and P.~R.~Page,
  %``Gluonic charmonium resonances at BaBar and Belle?,''
  Phys.\ Lett.\ B {\bf 628} (2005) 215  
  [arXiv:hep-ph/0507199].
  %%CITATION = HEP-PH 0507199;%%

%\cite{Pineda:2001zq}
\bibitem{Pineda:2001zq}
  A.~Pineda,
  %``Determination of the bottom quark mass from the Upsilon(1S) system,''
  JHEP {\bf 0106} (2001) 022
  [arXiv:hep-ph/0105008].
  %%CITATION = JHEPA,0106,022;%%

%\cite{Bali:2003jq}
\bibitem{Bali:2003jq}
  G.~S.~Bali and A.~Pineda,
  %``QCD phenomenology of static sources and gluonic excitations at short
  %distances,''
  Phys.\ Rev.\  D {\bf 69} (2004) 094001
  [arXiv:hep-ph/0310130].
  %%CITATION = PHRVA,D69,094001;%%

%\cite{Brambilla:2008zz}
\bibitem{Brambilla:2008zz}
  N.~Brambilla, A.~Vairo, A.~Polosa and J.~Soto,
  %``Round Table on Heavy Quarkonia and Exotic States,''
  Nucl.\ Phys.\ Proc.\ Suppl.\  {\bf 185} (2008) 107.
  %%CITATION = NUPHZ,185,107;%%

%\cite{Brambilla:2002nu}
\bibitem{Brambilla:2002nu}
  N.~Brambilla, D.~Eiras, A.~Pineda, J.~Soto and A.~Vairo,
  %``Inclusive decays of heavy quarkonium to light particles,''
  Phys.\ Rev.\  D {\bf 67} (2003) 034018
  [arXiv:hep-ph/0208019].
  %%CITATION = PHRVA,D67,034018;%%

%\cite{Tornqvist:1991ks}
\bibitem{Tornqvist:1991ks}
  N.~A.~Tornqvist,
  %``Possible large deuteron - like meson meson states bound by pions,''
  Phys.\ Rev.\ Lett.\  {\bf 67} (1991) 556.
  %%CITATION = PRLTA,67,556;%%

%\cite{Dubynskiy:2008mq}
\bibitem{Dubynskiy:2008mq}
  S.~Dubynskiy and M.~B.~Voloshin,
  %``Hadro-Charmonium,''
  Phys.\ Lett.\  B {\bf 666} (2008) 344
  [arXiv:0803.2224 [hep-ph]].
  %%CITATION = PHLTA,B666,344;%%

%\cite{Qiao:2005av}
\bibitem{Qiao:2005av}
  C.~F.~Qiao,
  %``One Explanation for the Exotic State Y(4260),''
  Phys.\ Lett.\  B {\bf 639} (2006) 263
  [arXiv:hep-ph/0510228].
  %%CITATION = PHLTA,B639,263;%%

%\cite{Jaffe:1976ig}
\bibitem{Jaffe:1976ig}
  R.~L.~Jaffe,
  %``Multi-Quark Hadrons. 1. The Phenomenology Of (2 Quark 2 Anti-Quark)
  %Mesons,''
  Phys.\ Rev.\  D {\bf 15} (1977) 267.
  %%CITATION = PHRVA,D15,267;%%

%\cite{Chistov09}
\bibitem{Chistov09}
  R.~Chistov, at this conference.

%\cite{Patrignani09}
\bibitem{Patrignani09}
  C.~Patrignani, at this conference.

%\cite{Choi:2003ue}
\bibitem{Choi:2003ue}
  S.~K.~Choi {\it et al.}  [BELLE Collaboration],
  % ``Observation of a new narrow charmonium state in exclusive B+- --> K+-  pi+
  %pi- J/psi decays,''
  Phys.\ Rev.\ Lett.\  {\bf 91} (2003) 262001 
  [arXiv:hep-ex/0309032].
  %%CITATION = HEP-EX 0309032;%%

%\cite{Aubert:2004ns}
\bibitem{Aubert:2004ns}
  B.~Aubert {\it et al.}  [BABAR Collaboration],
  % ``Study of the $B \to J/\psi K^- \pi^+ \pi^-$ decay and measurement of the $B
  %\to X(3872) K^-$ branching fraction,''
  Phys.\ Rev.\ D {\bf 71} (2005) 071103 
  [arXiv:hep-ex/0406022].
  %%CITATION = HEP-EX 0406022;%%

%\cite{Aubert:2008gu}
\bibitem{Aubert:2008gu}
  B.~Aubert {\it et al.}  [BABAR Collaboration],
  %``A Study of $B \to X(3872) K$, with $X_{3872} \to J/\Psi \pi^{+} \pi^{-}$,''
  Phys.\ Rev.\  D {\bf 77} (2008) 111101
  [arXiv:0803.2838 [hep-ex]].
  %%CITATION = PHRVA,D77,111101;%%

%\cite{Acosta:2003zx}
\bibitem{Acosta:2003zx}
  D.~Acosta {\it et al.}  [CDF II Collaboration],
  % ``Observation of the narrow state $X(3872) \to J/\psi \pi^+ \pi^-$ in
  %$\bar{p}p$  collisions at $\sqrt{s} = 1.96$ TeV,''
  Phys.\ Rev.\ Lett.\  {\bf 93} (2004) 072001 
  [arXiv:hep-ex/0312021];
  %%CITATION = HEP-EX 0312021;%%
  G.~Bauer, at the QWG meeting 2003.

%\cite{Abazov:2004kp}
\bibitem{Abazov:2004kp}
  V.~M.~Abazov {\it et al.}  [D0 Collaboration],
  % ``Observation and properties of the $X(3872)$ decaying to $J/\psi \pi^+
  %\pi^-$ in $p\bar{p}$ collisions at $\sqrt{s} = 1.96$ TeV,''
  Phys.\ Rev.\ Lett.\  {\bf 93} (2004) 162002 
  [arXiv:hep-ex/0405004].
  %%CITATION = HEP-EX 0405004;%%

%\cite{Aubert:2007rva}
\bibitem{Aubert:2007rva}
  B.~Aubert {\it et al.}  [BABAR Collaboration],
  %``Study of Resonances in Exclusive B Decays to Dbar(*)D(*)K,''
  Phys.\ Rev.\  D {\bf 77} (2008) 011102
  [arXiv:0708.1565 [hep-ex]].
  %%CITATION = PHRVA,D77,011102;%%

%\cite{Gokhroo:2006bt}
\bibitem{Gokhroo:2006bt}
  G.~Gokhroo {\it et al.},
  %``Observation of a near-threshold D0 anti-D0 pi0 enhancement in B --> D0
  %anti-D0 pi0 K decay,''
  Phys.\ Rev.\ Lett.\  {\bf 97} (2006) 162002
  [arXiv:hep-ex/0606055].
  %%CITATION = PRLTA,97,162002;%%

%\cite{Abe:2005ix}
\bibitem{Abe:2005ix}
  K.~Abe {\it et al.},
  % ``Evidence for X(3872) --> gamma J/psi and the sub-threshold decay X(3872)
  %--> omega J/psi,''
  arXiv:hep-ex/0505037.
  %%CITATION = HEP-EX 0505037;%%

%\cite{Abe:2005iy}
\bibitem{Abe:2005iy}
  K.~Abe {\it et al.},
  % ``Experimental constraints on the possible J(PC) quantum numbers of the
  %X(3872),''
  arXiv:hep-ex/0505038.
  %%CITATION = HEP-EX 0505038;%%

%\cite{Kravchenko:2006qx}
\bibitem{Kravchenko:2006qx}
  I.~Kravchenko  [CDF Collaboration],
  %``B spectroscopy at Tevatron,''
  eConf {\bf C060409}, 016 (2006) 
  [arXiv:hep-ex/0605\-076];
  %%CITATION = HEP-EX 0605076;%%
%\cite{Abulencia:2006ma}
%\bibitem{Abulencia:2006ma}
  A.~Abulencia {\it et al.}  [CDF Collaboration],
  %``Analysis of the quantum numbers J(PC) of the X(3872),''
  Phys.\ Rev.\ Lett.\  {\bf 98} (2007) 132002
  [arXiv:hep-ex/0612053].
  %%CITATION = PRLTA,98,132002;%%

%\cite{Voloshin:2006wf}
\bibitem{Voloshin:2006wf}
  M.~B.~Voloshin,
  %``Molecular quarkonium,''
  eConf {\bf C060409}, 014 (2006)
  [arXiv:hep-ph/0605063].
  %%CITATION = HEP-PH 0605063;%%

%\cite{Alexandrou:2006cq}
\bibitem{Alexandrou:2006cq}
  C.~Alexandrou, Ph.~de Forcrand and B.~Lucini,
  %``Evidence for diquarks in lattice QCD,''
  Phys.\ Rev.\ Lett.\  {\bf 97} (2006) 222002
  [arXiv:hep-lat/0609004];
  %%CITATION = PRLTA,97,222002;%%
%\cite{Fodor:2005qx}
%\bibitem{Fodor:2005qx}
  Z.~Fodor, C.~Hoelbling, M.~Mechtel and K.~Szabo,
  %``Nonperturbative investigation of the diquark potential,''
  PoS {\bf LAT2005} (2006) 310
  [arXiv:hep-lat/0511032].
  %%CITATION = POSCI,LAT2005,310;%%

%\cite{Hogaasen:2005jv}
\bibitem{Hogaasen:2005jv}
  H.~H\o gaasen, J.~M.~Richard and P.~Sorba,
  %``A chromomagnetic mechanism for the X(3872) resonance,''
  Phys.\ Rev.\ D {\bf 73} (2006) 054013 
  [arXiv:hep-ph/0511039];
  %%CITATION = HEP-PH 0511039;%%
%\cite{Buccella:2006fn}
%\bibitem{Buccella:2006fn}
  F.~Buccella, H.~Hogaasen, J.~M.~Richard and P.~Sorba,
  %``Chromomagnetism, flavour symmetry breaking and S-wave tetraquarks,''
  Eur.\ Phys.\ J.\  C {\bf 49} (2007) 743
  [arXiv:hep-ph/0608001].
  %%CITATION = EPHJA,C49,743;%%

%\cite{Maiani:2004vq}
\bibitem{Maiani:2004vq}
  L.~Maiani, F.~Piccinini, A.~D.~Polosa and V.~Riquer,
  %``Diquark-antidiquarks with hidden or open charm and the nature of
  %X(3872),''
  Phys.\ Rev.\  D {\bf 71} (2005) 014028
  [arXiv:hep-ph/0412098].
  %%CITATION = PHRVA,D71,014028;%%

%\cite{Brambilla:2005yk}
\bibitem{Brambilla:2005yk}
  N.~Brambilla, A.~Vairo and T.~R\"osch,
  %``Effective field theory Lagrangians for baryons with two and three heavy
  %quarks,''
  Phys.\ Rev.\  D {\bf 72} (2005) 034021
  [arXiv:hep-ph/0506065].
  %%CITATION = PHRVA,D72,034021;%%

%\cite{Maiani:2007vr}
\bibitem{Maiani:2007vr}
  L.~Maiani, A.~D.~Polosa and V.~Riquer,
  %``Indications of a Four-Quark Structure for the X(3872) and X(3876) Particles
  %from Recent Belle and BABAR Data,''
  Phys.\ Rev.\ Lett.\  {\bf 99} (2007) 182003
  [arXiv:0707.3354 [hep-ph]].
  %%CITATION = PRLTA,99,182003;%%

%\cite{Tornqvist:1993ng}
\bibitem{Tornqvist:1993ng}
  N.~A.~T\"ornqvist,
  % ``From the deuteron to deusons, an analysis of deuteron - like meson meson
  %bound states,''
  Z.\ Phys.\ C {\bf 61} (1994) 525 
  [arXiv:hep-ph/9310247];
  %%CITATION = HEP-PH 9310247;%%
%\cite{Tornqvist:2003na}
%\bibitem{Tornqvist:2003na}
  %N.~A.~Tornqvist,
  % ``Comment on the narrow charmonium state of BELLE at 3871.8-MeV as a
  %deuson,''
  arXiv:hep-ph/0308277.
  %%CITATION = HEP-PH 0308277;%%

%\cite{Swanson:2003tb}
\bibitem{Swanson:2003tb}
  E.~S.~Swanson,
  %``Short range structure in the X(3872),''
  Phys.\ Lett.\ B {\bf 588} (2004) 189 
  [arXiv:hep-ph/0311229];
  %%CITATION = HEP-PH 0311229;%%
%\cite{Swanson:2004pp}
%\bibitem{Swanson:2004pp}
%  E.~S.~Swanson,
  %``Diagnostic decays of the X(3872),''
  % Phys.\ Lett.\ B
  {\it ibid.} {\bf 598} (2004) 197 
  [arXiv:hep-ph/0406080].
  %%CITATION = HEP-PH 0406080;%%

%\cite{Pakvasa:2003ea}
\bibitem{Pakvasa:2003ea}
  S.~Pakvasa and M.~Suzuki,
  %``On the hidden charm state at 3872-MeV,''
  Phys.\ Lett.\ B {\bf 579} (2004) 67 
  [arXiv:hep-ph/0309294].
  %%CITATION = HEP-PH 0309294;%%

%\cite{Voloshin:2003nt}
\bibitem{Voloshin:2003nt}
  M.~B.~Voloshin,
  % ``Interference and binding effects in decays of possible molecular  component
  %of X(3872),''
  Phys.\ Lett.\ B {\bf 579} (2004) 316 
  [arXiv:hep-ph/0309307];
  %%CITATION = HEP-PH 0309307;%%
%\cite{Voloshin:2004mh}
%\bibitem{Voloshin:2004mh}
  %M.~B.~Voloshin,
  %``Heavy quark spin selection rule and the properties of the X(3872),''
  %Phys.\ Lett.\ B 
  {\it ibid.} {\bf 604} (2004) 69 
  [arXiv:hep-ph/0408321].
  %%CITATION = HEP-PH 0408321;%%

%\cite{Braaten:2003he}
\bibitem{Braaten:2003he}
  E.~Braaten and M.~Kusunoki,
  %``Low-energy universality and the new charmonium resonance at 3870-MeV,''
  Phys.\ Rev.\ D {\bf 69} (2004) 074005 
  [arXiv:hep-ph/0311147];
  %%CITATION = HEP-PH 0311147;%%
%\cite{Braaten:2005jj}
%\bibitem{Braaten:2005jj}
%  E.~Braaten and M.~Kusunoki,
  %``Factorization in the production and decay of the X(3872),''
  %Phys.\ Rev.\ D 
  {\it ibid.} {\bf 72} (2005) 014012 
  [arXiv:hep-ph/0506087].
  %%CITATION = HEP-PH 0506087;%%

%\cite{Cawlfield:2007dw}
\bibitem{Cawlfield:2007dw}
  C.~Cawlfield {\it et al.}  [CLEO Collaboration],
  %``A precision determination of the D0 mass,''
  Phys.\ Rev.\ Lett.\  {\bf 98} (2007) 092002
  [arXiv:hep-ex/0701016].
  %%CITATION = PRLTA,98,092002;%%

%\cite{Hanhart:2007yq}
\bibitem{Hanhart:2007yq}
  C.~Hanhart, Yu.~S.~Kalashnikova, A.~E.~Kudryavtsev and A.~V.~Nefediev,
  %``Reconciling the X(3872) with the near-threshold enhancement in the
  %D^0\bar{D}^{*0} final state,''
  Phys.\ Rev.\  D {\bf 76} (2007) 034007
  [arXiv:0704.0605 [hep-ph]].
  %%CITATION = PHRVA,D76,034007;%%

%\cite{Bignamini:2009sk}
\bibitem{Bignamini:2009sk}
  C.~Bignamini, B.~Grinstein, F.~Piccinini, A.~D.~Polosa and C.~Sabelli,
  %``Is the X(3872) Production Cross Section at Tevatron Compatible with a
  %Hadron Molecule Interpretation?,''
  Phys.\ Rev.\ Lett.\  {\bf 103} (2009) 162001
  [arXiv:0906.0882 [hep-ph]].
  %%CITATION = PRLTA,103,162001;%%

%\cite{Artoisenet:2009wk}
\bibitem{Artoisenet:2009wk}
  P.~Artoisenet and E.~Braaten,
  %``Production of the X(3872) at the Tevatron and the LHC,''
  arXiv:0911.2016 [hep-ph].
  %%CITATION = ARXIV:0911.2016;%%

%\cite{Eichten:1978tg}
\bibitem{Eichten:1978tg}
  E.~Eichten, K.~Gottfried, T.~Kinoshita, K.~D.~Lane and T.~M.~Yan,
  %``Charmonium: The Model,''
  Phys.\ Rev.\  D {\bf 17} (1978) 3090
  [Erratum-ibid.\  D {\bf 21} (1980) 313];
  %%CITATION = PHRVA,D17,3090;%%
%\cite{Eichten:2004uh}
%\bibitem{Eichten:2004uh}
  E.~J.~Eichten, K.~Lane and C.~Quigg,
  %``Charmonium levels near threshold and the narrow state $X(3872) \to
  %\pi^{+}\pi^{-}J/\psi$,''
  Phys.\ Rev.\  D {\bf 69} (2004) 094019
  [arXiv:hep-ph/0401210];
  %%CITATION = PHRVA,D69,094019;%%
%\cite{Eichten:2005ga}
%\bibitem{Eichten:2005ga}
  E.~J.~Eichten, K.~Lane and C.~Quigg,
  %``New states above charm threshold,''
  Phys.\ Rev.\  D {\bf 73} (2006) 014014
  [Erratum-ibid.\  D {\bf 73} (2006) 079903]
  [arXiv:hep-ph/0511179].
  %%CITATION = PHRVA,D73,014014;%%

%\cite{Le Yaouanc:1972ae}
\bibitem{Le Yaouanc:1972ae}
  A.~Le Yaouanc, L.~Oliver, O.~Pene and J.~C.~Raynal,
  %``Naive quark pair creation model of strong interaction vertices,''
  Phys.\ Rev.\  D {\bf 8} (1973) 2223;
  %%CITATION = PHRVA,D8,2223;%%
%\cite{Kalashnikova:2005ui}
%\bibitem{Kalashnikova:2005ui}
  Yu.~S.~Kalashnikova,
  %``Coupled-channel model for charmonium levels and an option for X(3872),''
  Phys.\ Rev.\  D {\bf 72} (2005) 034010
  [arXiv:hep-ph/0506270].
  %%CITATION = PHRVA,D72,034010;%%

%\cite{Bali:2005fu}
\bibitem{Bali:2005fu}
  G.~S.~Bali, H.~Neff, T.~D\"ussel, T.~Lippert and K.~Schilling  [SESAM
                  Collaboration],
  %``Observation of string breaking in QCD,''
  Phys.\ Rev.\  D {\bf 71} (2005) 114513
  [arXiv:hep-lat/0505012].
  %%CITATION = PHRVA,D71,114513;%%

%\cite{Bali09}
\bibitem{Bali09}
  G.S.~Bali, at this conference.

\end{thebibliography}
\end{document}